\RequirePackage{fix-cm}
\documentclass[smallextended]{svjour3}       
\smartqed  
\usepackage{graphicx}
\begin{document}

\title{MDia and POTS
}
\subtitle{The Munich Difference Imaging Analysis for the pre-OmegaTranS Project}


\author{Johannes Koppenhoefer, Roberto P. Saglia and Arno Riffeser
}


\institute{Max-Planck Institute for Extraterrestrial Physics \at
           Giessenbachstra\ss e, D-85741 Garching, Germany \\
           Tel.: +49-89-300003916\\
           Fax:  +49-89-300003495\\
           \email{saglia@mpe.mpg.de}
           \and
           Universit\"atssternwarte \at
           Scheinerstra\ss e 1, D-81679 M\"unchen, Germany \\
           Tel.: +49-89-21805995\\
           Fax:  +49-89-21806003\\
           \email{koppenh@usm.lmu.de,arri@usm.lmu.de}
}

\date{Received: 15 July 2011 / Accepted: 30 November 2011}

\maketitle

\begin{abstract}

  We describe the Munich Difference Imaging Analysis pipeline
  \linebreak that we developed and implemented in the framework of the
  Astro-WISE\footnote{Astronomical Wide-field Imaging System for
    Europe} package to automatically measure high precision light
  curves of a large number of stellar objects using the difference
  imaging approach. Combined with programs to detect time variability,
  this software can be used to search for planetary systems or binary
  stars with the transit method and for variable stars of different
  kinds. As a first scientific application, we discuss the data
  reduction and analysis performed with Astro-WISE on the
  pre-OmegaTranS data set, that we collected during a monitoring
  campaign of a dense stellar field with the Wide Field Imager at the
  ESO 2.2\,m telescope.

\keywords{databases astronomy \and extra-solar planets}
\end{abstract}
\section{Introduction}
\label{sec:Intro}
The Munich Difference Imaging Analysis Pipeline (MDia) is a software
tool to perform high precision photometric measurements in crowded
fields such as the Milky Way disk or the bulge of M31. MDia is based
on the difference imaging technique proposed by \cite{Tomaney96} and
\cite{Alard98}. Common applications are searches for transiting
extra-solar planets, microlensing surveys, stellar rotation and
oscillation studies as well as supernova projects.\\
The MDia tool convolves (degrades) a reference image (constructed from
the images collected during the periods of best seeing of the
observing campaign, see section
\ref{sec:MDIA_ReferenceFrameImplementation}) to match the seeing of
each single image in the data set. Subtracting the convolved reference
image from a single image, one obtains a difference image in which
every constant source disappears, while variable objects are visible
as positive or negative residuals. PSF-photometry performed on the
reference image and on each difference image at the position of the
detected sources allows to reconstruct the light curves of each
object.\\
\section{The Munich Difference Imaging Analysis Pipeline}
\label{sec:MDIA}
\subsection{The Algorithm}
\label{sec:MDIA_Algorithm}
As proposed by \cite{Alard98}, the optimal convolution kernel can be
modeled as the superposition of a set of Gaussian kernel base
functions which are modulated by a two-dimensional polynomial function
in the x- and y-direction:\\
\begin{equation}
	K(x,y) = \sum \limits_{i=1}^N \exp{\Large[-\frac{x^2+y^2}{2\sigma_i^2}\Large]} \sum \limits_{j=0}^{p_i} \sum \limits_{k=0}^{p_i-j} a_{ijk} x^jy^k \quad .
	\label{eq.kernel_base}
\end{equation}\\
The standard kernel model uses $N$=4 kernel base functions with
$\sigma_i$=\{0.1,9,6,3\} and $p_i$=\{0,2,4,6\} and a kernel size of
25\,x\,25 pixel. The total number of coefficients $a_{ijk}$ for this
kernel model is 50. Background differences are accounted for using a
polynomial model $B(x,y)$.\\
The coefficients $a_{ijk}$ and the background model are determined
simultaneously by $\chi^2$-minimization of the following
expression:\\
\begin{equation}
	\chi^2 = \sum \limits_{x,y} \frac{1}{\sigma^2(x,y)} [(R \otimes K)(x,y) + B(x,y) - S(x,y)]^2 \quad ,
	\label{eq.chi2}
\end{equation}\\
where $R$ is the reference image and $S$ is the single image.
Subtracting the convolved image $C := R \otimes K$ from each single 
image we obtain a difference image $D$:\\
\begin{equation}
	D(x,y) = \frac{S(x,y) - C(x,y)}{|K|} \quad ,
	\label{eq.difference}
\end{equation}\\
with\\
\begin{equation}
	|K| = \sum \limits_{x,y} K(x,y)
	\label{eq.norm}
\end{equation}\\
being the norm of the kernel. Using this normalization, the difference
image has the same flux level as the reference image. Note that in our
implementation we use a normalized kernel (i.e. $|K|$=1) because we
photometrically align all images before calculating the kernel ({\it
  skycalc} step, see below). In this case, any variable source could
degrade the quality of the kernel. To avoid this, the user can
optionally provide a mask that has zero values for each pixel that
should be excluded in the determination of the kernel. The
implementation of the above equations is described in \cite{Goessl}.\\
\subsection{The Astro-Wise Implementation}
\label{sec:MDIA_AWImplementation}
The Astro-WISE tool MDia is designed to operate on a set of regridded
images and associated weight images. The regridded images are fully
reduced images (bias-subtracted, flatfielded, etc.) that have been
resampled to a common target grid with a fixed pixel scale.\\
In the beginning, the user selects the best seeing images that will be
stacked to create a reference image. The second step is to create a
difference image for each input regridded image and to create the
light curves of all sources detected in the reference image. In the
next two sections we describe in detail the steps to create a
reference image and to create the light curves.\\
\subsubsection{Creating a Reference Image}
\label{sec:MDIA_ReferenceFrameImplementation}
\begin{figure}
\centering
\includegraphics[width=0.95\textwidth]{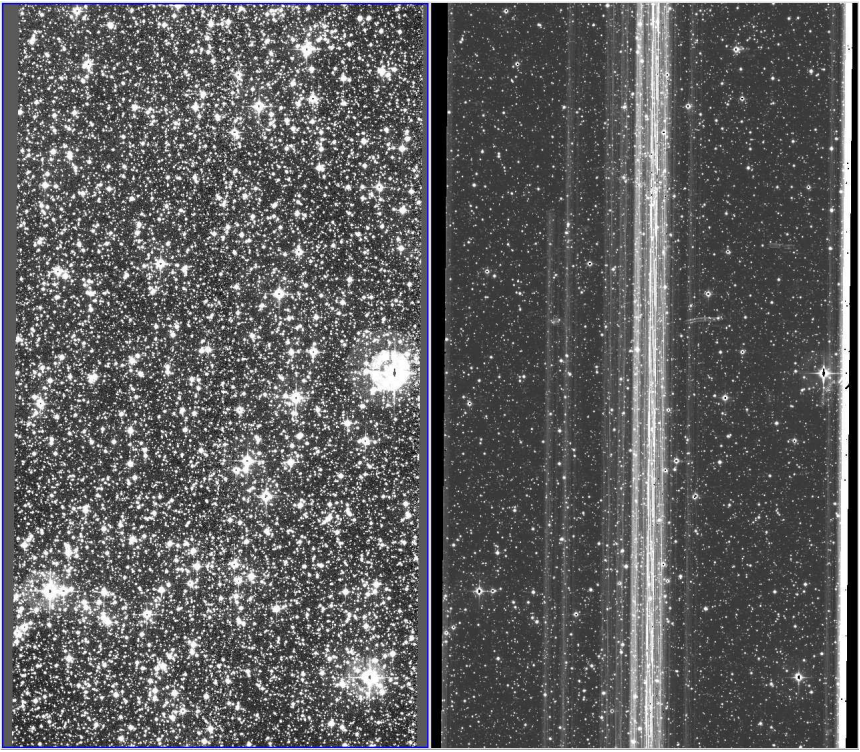}
\caption{Example reference image and its corresponding error
  image. Due to dithering masked pixels are on different x- and
  y-positions on the input regridded images and all bad pixel regions
  (e.g. bad columns) are therefore removed during the stacking
  process. Nevertheless, bad pixel regions have a larger variance,
  visible as brighter areas in the error image.}
\label{fig.ref}
\end{figure}

We recommend to use 30 input regridded images or more to create a
\linebreak reference image. With this number, masked areas in one
image can be replaced with the values of another image with similar
PSF. Only if the full data set does not contain 30 good seeing images
a lower number might be better because otherwise the input images will
have very different PSFs ({\it subby} step, see below). For code
examples and a complete list of all parameters and their meaning we
refer to the MDia
manual\footnote{http://www.usm.uni-muenchen.de/$\sim$koppenh/MDia}.\\\\
%
%
In the following we describe the processing steps that are performed
to create a reference image:\\
The first program that is executed is {\it prepare}. For each input
\linebreak regridded image it creates an associated error image using
the weight image and the read noise value. The error images (which are
inverse variance maps) are required by our code to perform a pixel by
pixel Gaussian error propagation. In a second step the program {\it
  wcscut} expands all input regridded images to cover a common area on
the sky which contains all pixels of the input images. After that, a
polynomial background is fitted and subtracted from the first input
image. To do this, the program {\it getsky} uses an iterative clipping
procedure to mask stars in the fit. In order to subtract the
background in all other images and to ensure that all images have the
same flux level, the program {\it skycalc} photometrically aligns each
input image to the background subtracted first input image.\\
The resulting reference image is a weighted stack of all input
\linebreak regridded images. The program {\it weight} calculates a
weighting factor for each image based on the background noise and the
PSF-FWHM as measured by SExtractor during the ingestion in
Astro-WISE. Before stacking, the program {\it subby} replaces masked
pixels in each input image with the pixel values of the other image
which has the most similar PSF. The similarity of the PSFs is
determined using a set of isolated stars. If a pixel is masked also in
the most similar image, {\it subby} replaces with the second most 
similar image, and so on.\\\\
The program {\it usmphot} optionally performs PSF-photometry on the
reference image in order to measure the fluxes of all sources. These
fluxes will be added at a later stage to the fluxes measured in the
difference images in order to create the light
curves. The program also outputs a PSF image.\\
In another optional step, the program {\it diffima} creates a set of
kernel base images that are stored and used for difference imaging
later.\\\\
Figure \ref{fig.ref} shows a reference image and the attached error
image that is used internally in our code. The reference image was
made using 30 input \linebreak regridded images with $\sim$0.7 arcsec
seeing.\\
\subsubsection{Creating the Light Curves}
\label{sec:MDIA_LightCurveImplementation}
\begin{figure}
  \includegraphics[width=0.95\textwidth]{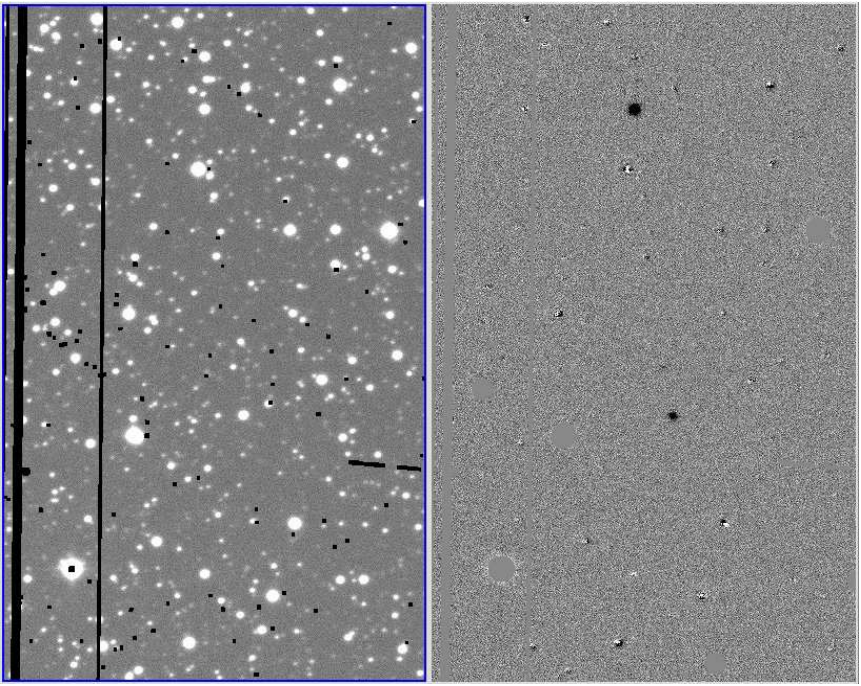}
  \caption{Example regridded image and the corresponding difference
    image. Two variable sources are visible as negative PSF-shaped
    residuals (dark spots). All constant sources are subtracted
    leaving only noise in the difference image.}
\label{fig.diff}
\end{figure}

Two inputs are required for the creation of light curves with MDia: A
reference image and a set of regridded images. For each input
regridded image, a difference image is created. PSF-photometry on the
difference image provides high precision measurements of the
differential flux of each object. Adding the constant flux of each
source as measured in the reference image (see previous section) the
light curves are created and stored as ASCII tables.\\
%
In practice the user splits the input regridded images in subsets of
typically 20 images and runs one job per subset. In this way the
processing is parallelized making use of a multi-node computer
cluster.\\\\
%
%
Figure \ref{fig.diff} shows an example input regridded image together
with the \linebreak difference image created with MDia. Two variable
sources are clearly visible as PSF-shaped negative residuals, meaning
that the flux in this particular regridded image was lower than in the
reference image.\\\\
\noindent In the following we describe the processing steps that are
performed to create light curves with MDia:\\
In the beginning the program {\it wcscut} expands and/or cuts the
input \linebreak regridded images to cover the same area on the sky as
the reference image\footnote{the input images may only have integer
  pixel shifts}. In the second step the program {\it skycalc}
photometrically aligns each image to the reference image, i.e. scales
and background corrects each image. The following step is the core of
the MDia tool. The program {\it diffima} uses the algorithms presented
in section \ref{sec:MDIA_Algorithm} to produce a difference image for
each regridded image. After that, the program {\it curvemaker}
constructs a PSF for each input image by combining a set of isolated
stars in the convolved reference image. These PSFs are then used to
perform PSF-photometry on the difference images and subsequently
create the light curves which are stored in ASCII format.\\
%
%
%
\subsection{Performance}
\label{sec:MDIA_performance}
We study the performance of MDia using a set of 1000 regridded images
which have been obtained for the pre-OmegaTranS project (see section
\ref{sec:POTS}). We limit our test to one chip (ccd50) of the ESO Wide
Field Imager camera. The size of the images is 2k\,x\,4k pixels. We
run our jobs on a Linux cluster which has 68 nodes. Each node is
equipped with two 2.6\,GHz dual-core AMD Opteron Processors, 6\,GByte
RAM and 200\,GByte local storage. We run only 2 jobs per node because
the processing requires a large amount of resources (i.e. memory, I/O
and disk space) and running more than two jobs simultaneously on one 
node turned out to be inefficient.\\
We execute each job multiple times and give the average execution time
in the following. In the first test we create a reference image using
20, 30 and 40 images without creating the kernel base images and
without measuring the reference image fluxes. The execution time is
68.5\,min, 84.2\,min and 144.5\,min respectively and therefore
roughly 3\,min per input image.\\
Including the creation of the kernel base images costs additional
15.9\,min, independent of the number of input images. Turning on
PSF-photometry in the reference image costs another 62.3\,min. We
recommend to include these optional steps when creating a reference
image because otherwise both will have to be executed each time the
reference image is used in MDia and the overall computation time will
be much higher.\\
In the second step we measure the performance of the MDia task. We use
the default kernel model with a kernel size of 25\,x\,25 pixel and
4\,x\,8 subfields. We measure the execution time of several jobs with
different number of input regridded images. For each job there is an
overhead contribution of 21.0\,min which is independent of the number
of images (retrieving the kernel base images, creating source lists,
storing logfiles, etc.). In addition to this overhead we find that
each regridded image needs about 5.7\,min. If we split e.g. all 1000
images in subsets of 20, the total execution time of all 50 jobs is
112.5\,h with an overhead contribution of 15.6\%.\\
\section{The Pre-OmegaTranS Project}
\label{sec:POTS}
\begin{figure}
  \centering
  \includegraphics[width=0.80\textwidth]{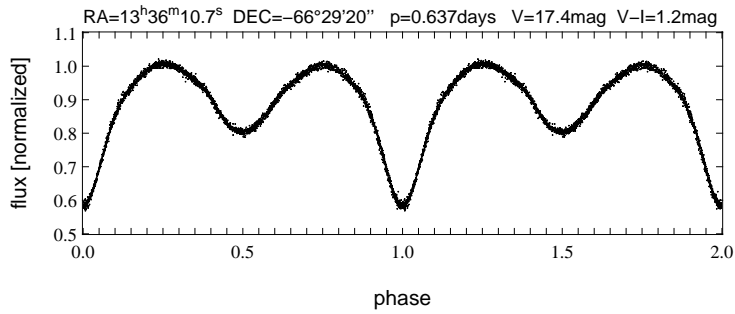}
  \caption{A $\beta$ Lyr type eclipsing binary. This close contact
    system consists of components with a different surface
    brightnesses. The light curve is characterized by strong
    ellipsoidal variations in between two eclipses of unequal depth.}
  \label{fig.EB1}
\end{figure}

\begin{figure}
  \centering
  \includegraphics[width=0.80\textwidth]{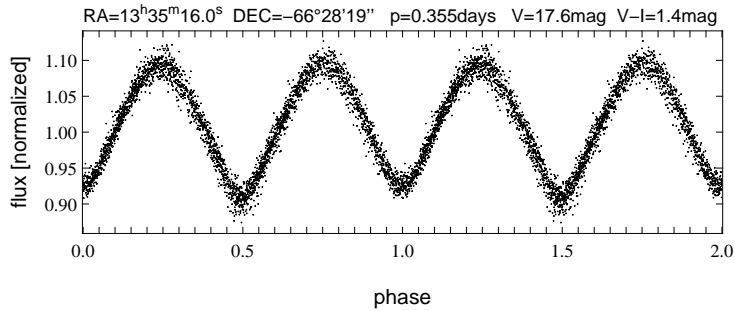}
  \caption{A non-eclipsing binary contact system with strong
    ellipsoidal variations at the 10\% level. In the pre-OmegaTranS
    light curve dataset we found $\sim$20 objects of this type.}
  \label{fig.EB2}
\end{figure}

\begin{figure}
  \centering
  \includegraphics[width=0.80\textwidth]{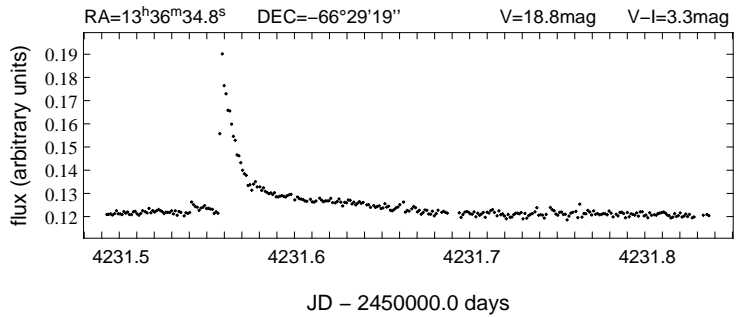}
  \caption{A UV Cet type flare star showing a single outburst on May
    11th 2007.}
  \label{fig.flare}
\end{figure}

In late 2004, a consortium of astronomers from INAF\footnote{Instituto
  Nazionale di Astrofisica} Capodimonte (Italy), Sterrewacht Leiden
(Netherlands) and MPE Garching (Germany) designed the OmegaCam Transit
Survey (OmegaTranS). A total of 26 nights of guaranteed time
observations with OmegaCam \cite{Kuijken} at the VLT Survey Telescope
\cite{Capaccioli} were granted to this project by the three
institutes. Scaling from existing surveys, OmegaTranS was expected to
deliver 10-15 new detections per year. Note that at that time only 8
transiting planets were known.\\
Due to delays in the construction and commissioning of the telescope,
the start of the project was delayed further and further and finally
canceled. Instead, we conducted a pre-OmegaTranS survey using the ESO
Wide Field Imager (WFI) mounted on the 2.2\,m telescope at LaSilla 
observatory \cite{Baade}.\\\\
As the outcome of 7 proposals (both for public ESO time and MPG
reserved time), a total of 129\,h of observations were collected in
the years 2006-2008. Spread over 34 nights we obtained a total of 4433
images of one field, i.e. OTSF-1a, in the Johnson R-band (filter
\#844, see WFI user manual). The image center of OTSF-1a is at
RA=13$^h$35$^m$41.6$^s$ and DEC=-66$^{\circ}$42'21'' (2000.0) and the
field dimensions are 34'\,x\,33'.\\
The exposure time was 25\,s in most cases. Under very good and very
bad observing conditions we slightly adjusted the exposure time in
order to achieve a stable S/N and to avoid saturating too many
stars. The average cycle rate (exposure, readout and file transfer
time) was 107\,s. 167 images with a seeing larger than 2.5\,arcsec
were not used because of their bad quality.\\
In addition to the science images, we obtained calibration images
(i.e. bias and flatfield exposures) for each of the 34 nights. The
total uncompressed raw data set comprises 725\,GBytes (589\,GBytes
science data, 136\,GBytes calibration data).\\
The basic CCD data reduction steps were done using the Astro-WISE
standard calibration pipeline. The steps include bias subtraction,
flatfield division, application of a bad pixel map, cosmic ray
filtering, astrometric calibration and remapping to a common target
pixel grid. For a complete description we refer to the Astro-WISE
manual.\\
Starting with regridded images we used the MDia tool to do
high-precision photometry as explained in section
\ref{sec:MDIA_AWImplementation}. We extracted the light curves of
16\,000 stars. At the bright end we reach an excellent photometric
precision of \mbox{2-3\,mmag} which is very close to the
theoretically expected precision.\\
Figs. \ref{fig.EB1} to \ref{fig.flare} show a few interesting example
light curves from the OmegaTranS data set. The transiting planet
candidates are presented in Koppenhoefer \mbox{et al.} (in prep.).\\
\section{Discussion and Conclusions}
\label{sec:Conclusions}
We described the Munich Difference Imaging Analysis pipeline. After
explaining the algorithm we presented the Astro-WISE implementation of
MDia. Since a complete description with code examples is beyond the
scope of this article we recommend any potential user to further read
the MDia manual).\\
In the last section we presented a subset of the results we obtained
with the MDia tool on the pre-OmegaTranS data set. We showed some
example light curves of variable stars that demonstrate the high
precision that can be achieved with the difference imaging technique.\\
%
%
%
%



\end{document}